\documentstyle[11pt,epsfig]{article}
\newcommand{\blankline}{\vskip .3cm}
\newcommand{\f}{\begin{equation}}
\newcommand{\ff}{\end{equation}}

\begin{document}
\centerline{\LARGE  Combinatorial dynamics in quantum gravity}
\blankline
\rm
\centerline{Stuart Kauffman${}^*$
and Lee Smolin${}^{**}$}
\blankline
\blankline
\centerline{\it ${}^*$  Biosgroup LTD}
\centerline{\it 317 Paseo de Peralta, Santa Fe, New Mexico, 
87501-8943}
\blankline
\centerline{${}^{**}$ \it  Center for Gravitational Physics and 
Geometry}
\centerline{\it Department of Physics}
 \centerline {\it The Pennsylvania State University}
\centerline{\it University Park, PA, USA 16802}
 \vfill
\centerline{July 20, 1998}
\vfill
\centerline{ABSTRACT}

We describe the application of methods from the study of
discrete dynamical systems to the study of histories of
evolving spin networks. These have been found to describe the
small scale structure of quantum general relativity and extensions
of them have been conjectured to give background independent formulations of
string theory. We explain why the
the usual equilibrium second order critical phenomena may not be
relevant for the problem of the continuum limit of such theories,
and why the relevant critical phenomena analogue to the problem of
the continuum limit is instead non-equilibrium critical phenomena
such as directed percolation.  The fact
that such non-equilibrium critical phenomena may be self-organized 
implies the possibility that the classical limit of quantum theories
of gravity may exist without fine tuning of parameters.

\blankline
email addresses:  stu@biosgroup.com, smolin@phys.psu.edu
\eject
\section{Introduction}

The idea that  space and time are fundamentally discrete is very old
and has often reappeared in the history of the search for a quantum
theory of gravity\footnote{see, for example \cite{roger-sn,finkelstein}.} 
However, it is only recently that concrete results
from attempts to construct a quantum theory have gravity have
been found which suggest very strongly that such a theory must
be based on a discrete structure.   These results come from
the quantization of general relativity\cite{vol1,sn1}, 
string theory\cite{string-dis}
and the thermodynamics of black holes\cite{bek,thooft,lenny-holo}.  
(For reviews 
see\cite{carlo-review,spain,future,garay}.)

If space and time are discrete, then the study of the 
dynamics of the spacetime may benefit from our understanding of other
discrete dynamical systems such as cellular automata\cite{wolfram}, 
froths\cite{froths} and binary networks\cite{binary}. The importance of this 
may be seen once it is
appreciated that  a key problem in any discrete theory of 
quantum gravity must be the recovery of continuous space time and
the fields that live on it as an approximation in an appropriate 
continuum limit.  This continuum limit, which will be also related
to the classical limit of the theory, (because the physical cutoff
$l_{Planck}$ which marks the transition between the discrete and
continuous picture is proportional to $\hbar$)
is then a problem in critical phenomena\cite{criticalcosmo}.
As one doesn't want the existence of classical spacetime to rest on 
some
fine tunings of parameters, this must presumably be some kind of
spontaneous, or self-organized critical 
phenomena\cite{soc}\footnote{Indeed this is a
general problem for particle physics, brought on by the hierarchy
problem, which is the existence of several widely separated scales.}.
 
However, there is a key element which 
which distinguishes quantum gravity from other
kinds of quantum and statistical 
systems This is that the causal structure is dynamical.   As a result,
the usual second order equilibrium critical phenomena may not be 
relevant for the continuum limit of quantum theories of gravity, as 
its connection
to quantum field theory relies on rotation from a Euclidean to 
Lorentzian metric and this is not well defined when the fluctuating 
degrees of freedom are the metric (or causal structure.)   
Instead, the relevant statistical physics analogue to the problem of
the classical limit will be non-equilibrium critical 
phenomena\cite{fl1}.
To see why, let us consider the issue of critical behavior for a 
discrete dynamical systems whose only attribute is causal structure.
Consider a set $\cal P$ 
of $N$ events, such that for any two of them $p$ and $q$
one may have either $p >q$, (meaning $p$ is to the causal future of 
$q$),
 or $q>p$, or neither, but never both.  This gives the set $\cal P$
the structure of a partially ordered set, or poset.  In addition, if
one assumes that there are no time like loops and that the poset
is locally finite (which means that there are only a finite number of
events in the intersection of the future of any event and the past
of any other) one has what is called a {\it causal set}.  One may then
invent an action which depend on the causal relations and
then study the quantum statistical physics of such a set, in the
limit of large $N$.   

This program has been pursued by physicists interested in using it
as a model of quantum gravity, particularly
by Myers, Sorkin\cite{rafael-poset}, `tHooft\cite{thooft-poset} and 
collaborators.  This is motivated by
the fact that the events of any Lorentzian spacetime form a
poset, where $p < q$ is the causal relation arising from the
lightcone structure of the metric.  In fact, if the causal
structure is given, the spactime metric is determined up to
an overall function.  

Sorkin and collaborators have conjectured that the causal structure 
is sufficient to define a satisfactory quantum 
theory of spacetime\cite{rafael-poset}.  
However, there is reason to believe that this may not be the case,
and that additional structure associated with what may loosely be
called the properties of space, must be introduced. One reason for 
this
is that the models where the degrees of freedom are only causal 
structure do not seem, at least so far, to have yielded the kinds of
results necessary to answer the key questions about the emergence
of the classical limit.

As a result, 
recently, Markopoulou proposed adding structure to poset models
of spacetime taken from results in other approaches to quantum gravity
\cite{fotini1}.   Her idea has been to combine the discrete causal 
structure 
of poset construction with descriptions of a discrete quantum
spatial geometry which has emerged from the study of non-perturbative 
quantum gravity.  These descriptions are usually expressed in terms 
of spin networks, which are graphs whose edges are
labeled with half-integers,
$1/2,1,3/2,...$ which represent quantum mechanical spins.  
Originally invented by Penrose\cite{roger-sn}, more recently they have been 
shown
to represent faithfully a basis of exact non-perturbative states of
the quantum gravitational field\cite{vol1,sn1}.  Extensions
of the spin network states have also been constructed that
are relevant for supergravity\cite{supersn} and other extensions
have been proposed in the context of a conjectured background
independent formulation of string theory\cite{tubes,pqtubes}

To show how the discrete causal structure of posets may be
fitted to a discrete description of both spacetime and 
spatial geometry we may 
need to describe the structure of a causal set  $\cal P$
in more detail.  The {\it Alexandrov neighborhood}
of two events $p$ and $q$, $A(p,q)$, consist of all $x$ such that
$p < x  < q$.  't Hooft has proposed that the number of events in
$A(p,q)$ should be a measure of its volume, in Planck units.
If the poset is taken by events picked randomly from a Lorentzian
manifold, using the measure given by the volume element, there
is then exactly enough information in the poset to reconstruct the
metric, in the limit of an infinite number of events.
Using the Alexandrov neighborhoods of a poset, we may then construct
a discrete model of a  spacetime geometry.  When the theory has a
good classical limit that should approximate a continuous spacetime
geometry.

In classical general relativity it is possible to define an infinite 
number 
of spatial slices, which have defined on them three dimensional
Reimannian geometries.  There are an infinite number of ways to
slice a spacetime into a sequence of spatial slices, each of which 
may be
associated with surfaces of simultaneity defined by a family of
observers and clocks moving in the spacetime.  Because the
choice of how to slice spacetime into a series of spatial geometries 
is arbitrary time in general relativity is referred to as being
``many-fingered".

A completely analogous 
notion of spatial geometry can be defined strictly in terms of a 
poset.
To do this we 
consider a set of events $\Sigma \subset {\cal P}$ which consists
of events $y_i$ such that no two of them are causally related
(i.e. neither $y_i < y_j$ or $y_j < y_i$ for all pairs in $\Sigma$.)
These may be called ``spacelike related".  If no event of 
$\cal P$ may be added to $\Sigma$ preserving the condition of
no causal relations it is a maximal set of spacelike related events.
Such sets are called {\it antichains} or {\it discrete spacelike 
slices}
of $\cal P$. 
The basic idea  of \cite{fotini1} is then to  endow the antichains of 
causal sets with the properties of discrete quantum geometries 
represented
by spin networks.  The result gives a notion of a quantum 
spacetime, which is discrete but which has many of the attributes 
of
continuous spacetime, including causal structure, spacelike slices 
and many-fingered time.  As described in \cite{fotini1} discrete sets having
these properties can be constructed by beginning with a spin network
and then altering it by a series of local moves.  

The purpose of this paper is to raise several key issues involved in
the study of the continuum limits in this kind of formulation of
quantum gravity.  It is written for
statistical physicists, relativists and quantum field 
theorists.  Our intention in writing it is mainly
to point the attention of people in these fields to the existence
of a class of problems in which methods used to study non-equilibrium
critical phenomena may play an important role in studies of
quantum gravity.  

In the next section we describe the basic structure of a causally
evolving spin network, in language we hope is 
accessible to statistical physicists.   We do not give any details
about how these structures are related to general relativity or
its quantization, these may be found 
elsewhere\cite{vol1,sn1,lp1,lp2,spain,carlo-review,future}.  Section
3 and 4 then discuss the problem of the classical limit of this theory
In section 5 some structures are defined on the set of quantum
states of the theory, which are then used in sections 6 and 7, 
in the context of a simplified model, to argue for the existence of
a classical limit that may reproduce general relativity.
Section 8 then introduces a new question, which is how the dynamics
of the theory is to be chosen.  We suggest that it may be reasonable
for the dynamics to evolve as the spacetime does, leading to the
classical limit as a kind of self-organized critical phenomena.

\section{Combinatorial descriptions of quantum spacetime}

There are actually several closely related versions of the spin
network description of quantum 
spatial geometry\cite{qdef,tubes,pqtubes}.  
As our interest
here is on the analysis of their dynamics, we will consider only one
kind of model, which is the easiest to visualize.  This is
associated with combinatorial 
triangulations\cite{fotini1}\footnote{Its exact 
relationship to the spin network states which arise in canonical 
quantum gravity is complicated, due to some subtleties which need not 
concern us here.  These are discussed in \cite{prep}.}.

We describe first the quantum geometry of space, then how these
evolve to make combinatorial spacetimes.

\subsection{Combinatorial description of spatial geometry}

A combinatorial $m$-simplex is a set of $m$ points, 
$e_1,...e_m$ called the vertices,
together with all the subsets of those points.  Those subsets with
two elements, $e_{12}=\{ e_1 , e_2 \}...$ are called edges, those
with three $e_{123}=\{ e_1 , e_2 , e_3 \}...$ faces and so on.
A combinatorial tetrahedron is a combinatorial $4$ simplex.

A three dimensional simplicial psuedomanifold, $T$, consists
of a set of $N$ combinatorial tetrahedra joined such that each
face is in exactly two tetrahedra.  Many such psuedomanifolds
define manifolds, in which case the neighborhoods of the edges and
nodes are homeomorphic to the neighborhoods of edges and nodes 
in triangulations of Euclidean three space.   These are constraints
on the construction of the psuedomanifold, which are called the
manifold conditions.  When they are not satisfied, we have a more
general structure of a psuedomanifold.  Many psuedomanifolds can
be constructed from manifolds by identifying two or more edges or
nodes.

The sets on which the manifold conditions fail to be satisfied
constitute defects in the topology defined by the combinatorial
triangulation.  Under suitable choices of the evolution rules these
defects propagate in time, forming extended objects, with dimension
up to two less than the dimension of the spacetime.  When the
discrete spacetime has a dynamics, as we will describe below,
laws of motion for the extended objects are induced.  It is very
interesting that string theory in its present form has in it extended
objects of various dimensions; the relationship between those 
``branes" and the defects in psuedomanifolds is under
investigation\cite{prep}.

A psuedomanifold may be labeled by attaching suitable labels
to the faces and tetrahedra.  For quantum gravity it is useful to 
consider
labels that come from the representation theory of some algebra
$G$, which may be a Lie algebra, a quantum Lie algebra, a 
supersymmetry algebra, or something more general.   Such
algebras are characterized by a set of representations, $i,j,k...$
and by product rules for decomposing products of representations,
$j \otimes k = \sum_l f_{jk}^l l$, where the $f_{jk}^l$ are integers.
Each such algebra has associated to it linear vector spaces
$V_{ijkl}$, which consists of the linear maps
$\mu : i \otimes j \otimes k \otimes l \rightarrow 1$, where
$1$ is the one dimensional identity representation.  It is then
usual to label a model of quantum gravity with algebra $G$
by associating a representation $k$ with each face and an
intertwinor $\mu \in V_{ijkl}$ to each tetrahedra, where
$i,j,k,l$ label its four faces.  The pseudomanifold $T$, together
with a set of labels is denoted $\Gamma$ and called a
labeled pseudomanifold.  

It is particularly convenient to work with a quantum group at
a root of unity, as the label sets in these cases are finite.  In
canonical quantum gravity, the
quantum deformation is related to the cosmological 
constant\cite{linking,hologr}.

To each labeled pseudomanifold $\Gamma$ we associate a
basis state $|\Gamma >$  of a quantum theory of gravity. The
set of such states spans the state space of the theory, 
$\cal H$, whose inner product is chosen so that the topologically
distinct $|\Gamma >$'s comprise an orthonormal basis.

Each labeled pseudomanifold is also dual to a {\it spin network},
which is a combinatorial graph constructed by drawing an
edge going through each face and joining the four edges that
enter every tetrahedra at a vertex\cite{roger-sn,sn1}.  
The edges are then labeled
by representations and the nodes by intertwinors\footnote{
Note that the pseudomanifolds have more information than
the spin networks, for a given spin network may come from
several combinatorial triangulations.  The spin network structure
may be extended so as to code this additional information, for
example by extending the edges into tubes as in 
\cite{tubes,pqtubes}.  For
simplicity in this paper we stick to psuedomanifolds.  In some
papers these are also called ``dual spin networks''\cite{fotini1}.}.

If one wants a simpler model one may simply declare all  labels
to be identical and leave them out.  These are called ``frozen
models\cite{prep}".  Frozen models are like the dynamical 
triangulation
models of Euclidean quantum gravity, except that there are different
kinds of simplices, corresponding to causal ordering.  We may
also consider ``partly frozen" models in which the spins on the
faces are all equal, but the intertwinors are allowed to vary over
a set of allowed values.

One of  the results of the canonical quantization of general 
relativity
is a geometrical interpretation for the spins and intertwinors of spin
networks.  Given the correspondence of labeled triangulations to 
spin networks, this interpretation may be applied directly to the
simplices of the labeled spin networks.  Doing this, we find that
each face $f_{abc}$ of the combinatorial triangulation has an
area, which is related to the spin $j_{abc}$ on the face by
the formula\cite{vol1},
\f
A_{abc}= l_{Planck}^2 \sqrt{j_{abc} ( j_{abc}+1 )}
\ff

There are also quanta of volume associated with the
combinatorial tetrahedras of the combinatorial 
triangulations.  This correspondence is more complicated,
and is motivated as well from canonical quantum gravity.
Associated with the finite dimensional 
space of intertwinors ${\cal H}_{j_\alpha}$
at each node, where the spins of the $4$ incident edges are
fixed to be $j_\alpha$, is a volume operator
${\cal V}_{j_\alpha}$\cite{spain,vol1}.  
These operators are constructed
in canonical quantum gravity\cite{spain,vol1} and shown to be
hermitian\cite{vol2}.  They are also finite
and diffeomorphism invariant, when constructed through
an appropriate regularization procedure\cite{spain,vol1}.
Their spectra have been computed\cite{vol2},
yielding a set of eigenvalues $\{ v^I_{j_\alpha} \}$
and eigenstates $|v^I_{j_\alpha}  > \in {\cal H}_{j_\alpha}$.
These eigenvalues are given, in units of $l_{Planck}^3$
by certain combinatorial expressions found in \cite{vol2}.
Thus, a combinatorial
triangulation represents a quantum geometry where
the faces have areas and the tetrahedra volumes, which
depend on the labelings in the way we have described.

\subsection{Causal evolution of quantum geometries}

We now follow the proposal of  \cite{fotini1}
and construct combinatorial quantum spacetimes by
applying a set of evolution rules to the states we
have just described.  A basis state 
$|\Gamma_0 > \in {\cal H}$ may evolve to one of 
a finite number of possible successor states $| \Gamma_0^I >$.
Each $|\Gamma^I_0 >$ is derived from $| \Gamma_0 >$ by
application of one of four possible moves, called
Pachner moves\cite{}.  These moves modify the state $| \Gamma_0 >$
in a local region involving one to four adjacent
tetrahedra.

Consider
any subset of $\Gamma$ consisting of $n$ adjacent tetrahedra, 
where $n$ is between $1$ and $4$, which make up $n$  out of the
$5$
tetrahedra of a 
four-simplex $S_4$.  Then there is an evolution rule by which those 
$n$
tetrahedra are removed, and replaced by the other $5-n$ tetrahedra
in the $S_4$.  This is called a Pachner move.  The different possible
moves are called $n \rightarrow (5-n)$ moves 
(Thus, there are $1\rightarrow 4$, 
$2 \rightarrow 3 $, etc. moves.  
The new tetrahedra must be labeled, by 
new representations $j$ and intertwiners $k$.  
For each move there are $15$ labels involved, 
$10$ representations on the faces 
and $5$ intertwinors on the tetrahedra.  This is because
the labels involved in the move are exactly those of the
four simplex $S_4$.  
For each $n$  there is then an amplitude
${\cal A}_{n \rightarrow 5-n}$ that is a function of the $15$ labels.  A
choice of these amplitudes for all possible labels, for the four 
cases $1\rightarrow 4 ,....,4\rightarrow 1$, then constitutes 
a choice of the dynamics
of the theory.

The application of one of the possible Pachner moves
to $\Gamma_0$, together with a choice of the possible
labelings on the new faces and tetrahedra the move
creates, results in a new labeled pseudomanifold state $\Gamma_1$.
This differs from $\Gamma_0$ just in a region which
consisted of between $1$ and $4$ adjacent tetrahedra.
The process may be continued a finite number of times
$N$, to yield successor labeled pseudomanifold states 
$\Gamma_2, ... \Gamma_N$.

Any particular set of $N$ moves beginning with a state
$\Gamma_0$ and ending with a state
$\Gamma_N$ defines a four dimensional 
combinatorial structure, which we will call a
{\it history}, $\cal M$ from $\Gamma_0$ to $\Gamma_N$.
Each history consists of $N$ combinatorial four simplices.
The boundary of $\cal M$,  is a set of tetrahedra which
fall into two connected sets so that
$\partial {\cal M} = \Gamma_0 \cup \Gamma_1$.  
All tetrahedra not in the boundary of $\cal M$ are
contained in exactly two four simplices of $\cal M$.

Each history $\cal M$ is a causal set, whose structure
is determined as follows.  The tetrahedra of each
four simplex, $S_4$ of $\cal M$ are divided into two
sets, which are called the past and the future set.
This is possible because each four simplex contains
tetrahedra in two states $\Gamma_i$ and $\Gamma_{i+1}$
for some $i$ between $0$ and $N$.
Those in $\Gamma_i$ were in the group that were wiped
out by the Pachner move, which were replaced by those
in $\Gamma_{i+1}$.  Those that were wiped out are
called the past set of that four simplex, the new ones,
those in $\Gamma_{i+1}$ are called the future set.
With the exception of those in the boundary, every
tetrahedron is in the future set of one four simplex
and the past set of another.

The causal structure of $\cal M$ is then defined as follows.
The tetrahedra of $\cal M$ make up a causal set
defined as follows.
Given two tetrahedra $T_1$ and $T_2$ in $\cal M$, we
say $T_2$ is to the future of $T_1$ (written $T_2 > T_1$)
iff there is a sequence of causal steps that begin on
$T_1$ and end on $T_2$.  A causal step is a step from
a tetrahedron which is an element of the past set of
some four simplex, $S_4$ to any tetrahedron which is
an element of the future set of the same four simplex.
By construction, there are no closed causal loops, so the
partial ordering gives a causal set.

Each history $\cal M$ may also be foliated by a number of
spacelike slices $\Gamma$.  These are the anitchains that
we defined in section 1

Each $\Gamma_i$ in the original construction of
$\cal M$ constitutes a spacelike slice of $\cal M$.
But there are also many other spacelike slices
in $\cal M$ that are not one of the $\Gamma_i$.
In fact, given any spacelike slice $\Gamma$ in
$\cal M$ there are a large, but finite, number of
slices which are differ from it by the application
of one Pachner move.  Because of this, there
is in this formulation a discrete analogue of the
many fingered time of the canonical picture
of general relativity.  

\subsection{How the dynamics are specified}

We have now defined quantum spatial geometry and quantum
spacetime histories, both completely combinatorially.   To turn this 
structure into a physical theory we must invent some dynamics.  
Although it is not the only possible starting point (and we will 
discuss another in section 8) it is best to begin by being 
conservative and using the standard notion of the path integral. We 
then assign to each
history $\cal M$ an amplitude ${\cal A}[{\cal M}]$ given by
\f
{\cal A}[{\cal M}] = \prod_i A[i]
\ff
where the product is over the moves, or equivalently the 4-simplices,
labeled by $i$. $A[i]$ is the amplitude for that four simplex, which
will be a function of its causal structure ($1\rightarrow 4$ or the
others) and the labels on its faces and tetrahedra.  The dynamics
is specified by giving the complex function $A[i]$, which depends on 
the possible causal structures and labels, a choice of such a 
function is 
equivalent to a choice of an action.

The amplitude for the transition from an initial state
$|i>$ to a final state $|f>$, both in $\cal H$ is then given by
\f
T[i,f] = \sum_{{\cal M}|{\partial {\cal M}}= |i> \cup |f>} 
{\cal A}[{\cal M}]
\label{pi}
\ff
 where the sum is over all histories from the given initial and
final state.

The theory is then specified by giving the kinematics, which is the
algebra from which the label set is chosen and the dynamics, which
is the choice of functions $A[i]$.  One important question, which we
will now discuss, is whether there are choices that lead to theories 
that have a good classical limit.

\section{The problem of the classical limit and its relationship
to critical phenomena}

Having defined the class of models we will study, we now turn to our 
main subject, which is the problem of the classical limit and its 
relation to
problems in non-equilibrium critical phenomena.   We begin by making 
the following observation:  Suppose that the amplitudes of each move 
were real numbers of the form,
\f
A[i] = e^{-S(i)}
\ff
Then the sum over histories can be considered to define a statistical 
system, whose partition function is of the form,
\f
Z[i,f] =  \sum_{{\cal M}|{\partial {\cal M}}= |i> \cup |f>} 
e^{-\sum_i S(i)}
\ff
Thus we have a statistical average over histories, each weighed by a 
probability, just as in non-equilibrium systems such as percolation 
problems.  In fact, there is an exact relationship with directed 
percolation problems, as the following example shows.

\begin{figure}
\centerline{\mbox{\epsfig{file=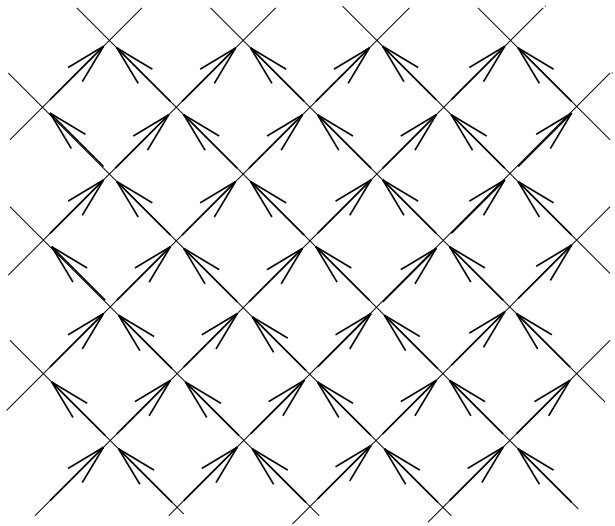}}}
\caption{A $1+1$ dimensional directed percolation problem.}
\label{setup}
\end{figure}

\begin{figure}
\centerline{\mbox{\epsfig{file=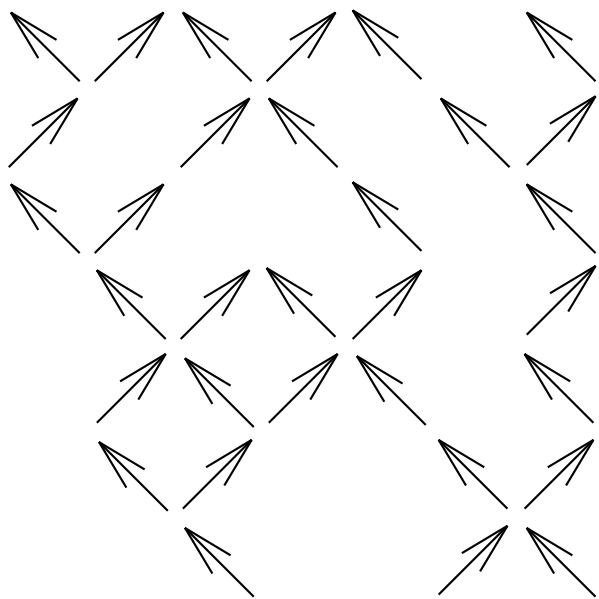}}}
\caption{One history of a directed percolation system.}
\label{hist}
\end{figure}

In Figure 
(\ref{setup}) 
we show the setup 
of a $1+1$ directed
percolation problem.   
The degrees of freedom are the arrows, each of which points
to the future, which is upwards in the picture.  The
{\it value} or state of an arrow is whether it is on or off.  A 
history,
$\cal M$ of a directed percolation problem is a record of
which arrows are on.  One such history is shown in Figure (\ref{hist}).

In the simplest version of directed
percolation, each arrow is turned on with a probability $p$.
There is a critical probability $p^*$ at which the percolation phase
transition takes place.  Below $p^*$ the on arrows make up
disconnected clusters of finite size, whereas for $p > p^*$ the
on arrows almost always form a single connected cluster.
At $p^*$ the system is just barely connected.  At this point
correlation functions are scale invariant.

A more complicated version of directed percolation can be
described as follows.
Each diagonal link is turned on or off according
to a rule which depends on several parameters.  To do this one
introduces a time coordinate, which is a label attached
to the nodes which is increasing in the
direction the arrows point and so that all nodes that share a
common time coordinate are causally unrelated.  We then 
apply the rule to each node at a given time, successively in
time, generating the evolution of the history from some initial
state.

Each node has two arrows pointing
towards it, which we will call the node's past arrows and two arrows
leaving it, which we will call its future arrows.  
The rule governs whether one or both of
the future pointing arrows at the node are on, as a function of the 
state of the past arrows.  For our purposes the exact
form of the rules is not important, what matters is that there is a 
critical
surface in the space of parameters at which the behavior of the system
is critical, corresponding to the percolation phase transition.  
At the critical point the system is in the same universality class as
simple directed percolation depending on the one parameter $p$.
This second model will be called the dynamical model, as 
the histories evolve in time, by applying the rule to the nodes at
later and later times.  A dynamical model may be probabilistic
or deterministic, depending on the nature of the rule applied at
each node.

Notice that a history $\cal M$ of a directed percolation problem is
a causal set.  We will say that a node $p$ is to the future of a
node $q$ (and write $p>q$) in a given history $\cal M$ 
if there is a chain of on arrows beginning
at $q$ and ending at $p$. A model of directed percolation in
$d+1$ dimensions is then a model of dynamical causal structure
for a discrete $d+1$ dimensional spacetime.  A  history $\cal M$
of a directed percolation model 
then has a causal structure and all its acutraments, including 
discrete spacelike
surfaces, light cones, future causal domains, past causal domains, 
etc.
In a percolation problem based on a fixed spacetime lattice
as in Figure (\ref{setup} we may define the {\it background causal
structure} to be the one defined by the history in which
all the arrows are on.

In particular, the values of the arrows (on or off) at one
time $t$ make a state $|\psi >$.  If the model has $n$  
arrows in each constant time surface, the 
state space is $4^n$ dimensional. In the deterministic
models an initial state $|\psi_0>$ evolves to a unique history
$\cal M$.   Thus a deterministic model of directed percolation is
a cellular automata, called 
a Domany-Kunsel cellular automata model\cite{DK}.

One way to understand what happens at the directed percolation 
critical
point is to use the concept of damage\cite{binary}.  In a deterministic
model of directed percolation pick an initial state
$|\psi_0>$.  Evolve the system to  a history ${\cal M}_0$.
Then change one arrow $a_0$ in the initial state and evolve to the
corresponding history ${\cal M}_1$.  Label any arrow whose
value is different in the two histories as {\it damaged}.  The
damaged arrows make a connected set $\cal D$, called
the damaged set, which lie
in the future causal domain of the arrow $a_0$ according to the
background causal structure.

Hence, we see that damage corresponds to a perturbation of the 
discrete
causal structure.  It is interesting to ask how the morphology of the
damaged region depends on the phase of the percolation system.
Below the percolation phase transition the causal domains are
finite and isolated, and the same is true for the damaged sets.
Just at the phase transition point, damage is able to propagate arbitrarily
far, for the first time.  However, the damage is constrained to follow
the background causal structure, which is the causal structure of
the unperturbed history.  Thus, if the theory has a continuum
limit, the spread of the damage will correspond
to the propagation of some causal effect.  But if there is a continuum
limit associated with the phase transition, then 
the correlation functions
that measure the spread of damage will be power-law. In this case
they should correspond in the continuum limit to the propagation
of massless particles.  Thus, 
if we think of the damage as the propagation of a perturbation in the
causal structure, it must correspond in the continuum limit to the
propagation of a graviton, which is how 
the propagation of a change in the causal structure
is described in the perturbative theory.  If the theory has a good
continuum limit then the gravitons must travel arbitrarily far
up the lightcones of the background causal structure.  We see that
this will only be possible at the critical point of the directed
percolation model.

Thus, by identifying a directed percolation model with a dynamical
theory of causal structure, we see that if that theory is to have
a continuum limit corresponding to general relativity in $4$ or
more spacetime dimensions, the only possibility for the existence of
such a limit is at the critical point of the directed percolation 
model.
Thus we see that directed percolation critical phenomena must play
the same role for discrete models of dynamical causal structure that
ordinary second order critical behavior plays in Euclidean quantum
field theory.

\section{Is there quantum directed percolation?}

There is however an important difference between what is required for
a theory of quantum gravity and the directed percolation models so
far studied by statistical physicists.  
In a discrete model of quantum gravity each history 
$\cal M$ is assigned an amplitude ${\cal A}[{\cal M}]$, which is
generally a complex number.  All directed percolation models so far 
studied (to the authors' knowledge)
are either deterministic or probabilistic.  In the latter case a
probability $p[{\cal M}]$ is assigned to each history $\cal M$,
which is of course a real number between $0$ and $1$.  
It is only in this case, in which each history has a probability,
that we know anything about the critical phenomena associated
to directed percolation.   
However, in quantum mechanics paths are weighed by amplitudes,which 
are complex numbers.  Thus, 
it would thus be very interesting to
know whether there are analogous critical phenomena in models
which are set up as directed percolation models, (for example
as in Figure (\ref{setup}), except that
a complex amplitude ${\cal A}[e]$,
rather than a probability, is assigned to
the state at each node.  We may call such a model a 
{\it quantum directed percolation model}.   
We believe that the study of such models could be very useful
for understanding the conditions required for discrete models
of quantum gravity to have good continuum limits.

One issue that must be stressed is that very little is actually known
about the continuum limit for Lorentzian path integrals where the
histories are weighed by complex phases rather than probabilities.
In quantum mechanics and conventional quantum field theory the path
integrals are normally {\it defined} by analytic continuation from  
Euclidean field theory, where the weights can be considered
probabilities.  In the absence of such a definition, one might try
to define the sum over histories directly.   However, one faces 
a serious question of whether the sums converge at all.

This problem cannot be avoided in a case such as the present, in which
the system is discrete.  Of course, the usual wisdom is that  
the classical limit
will exist because the phases from histories which are far-from-classical
paths interfere destructively, leaving only the contributions
near-classical histories, which add constructively.  The problem is
that in a finite system, in which there are a finite number of
histories in the sum, the cancellation coming from the destructive
interference will not be complete.  There will be a residue
coming from the sum, with a random phase and an
absolute value of order $\sqrt{n}$, if there
are $n$ far-from-classical histories\footnote{This has been verified
in a numerical computation by Sameer Gupta.}.  This contribution must
be much smaller than those coming from close to classical paths,
which will have an absolute value of order $m$, where $m$ is the
number of close to classical paths.  Thus, the existence of the
classical limit seems to require that $m >> \sqrt{n}$, which
means that there are many more near classical paths than
far-from-classical paths.  Of course, in any standard quantum
system the actual situation is the opposite, there are many
more far-from-classical than near-classical paths.

This argument suggests that the existence of the classical
limit may require that a continuum limit has been taken in which
the number of histories diverges.  In this case it may be
possible to tune parameters to define a limit in which the
non-classical contribution to the amplitude cancels completely.
In essence, this is what is forced by defining the theory in
terms of an analytic continuation from a Euclidean field theory.

In the absence of a definition by an analytic continuation,
the sums over causal histories
may fail to have a good classical limit because they lack
both an infinite sum over histories and a suitable definition of
a corresponding Euclidean theory.  This is perhaps the key
question concerning the classical limit of such theories.

\section{Discrete superspace and its structure}

Having raised several issues concerned with the evaluation of the
path integrals that arise in studies of evolving spin networks,
we would now like to describe here a formalism and a language which may
be useful for addressing them.  It is convenient
to consider a superspace $\Omega$ consisting of all 
$3$ dimensional psuedomanifolds constructed with a finite number
of tetrahedra.   Associated to this is $\Omega_G$, which is the
space of all labeled pseudomanifolds based on
the algebra $G$.  These spaces have
intrinsic structure generated by the evolution under the
Pachner moves.

Consider an initial  pseudomanifold $\Gamma^0$,
with a finite number of tetrahedra.  
We then consider all pseudomanifolds 
$\gamma^1_\alpha$ that can
be reached from $\Gamma^0$ by one instance of any of the 
$4$ allowed
moves $n \rightarrow 5-n $.  They are finite in
number, and labeled by an arbitrary integer $\alpha$.  We will
call this set ${\cal S}^1_{\gamma^0}$.   Generalizing this, it is
natural then to consider the set ${\cal S}^N_{\gamma^0}$ of
all pseudomanifolds that can be reached from $\Gamma^0$ in
$N$ or less moves.  Clearly we have 
${\cal S}^{N-1}_{\gamma^0} \subset {\cal S}^N_{\gamma^0}$.
We will also want to speak about the ``boundary" of
${\cal S}^N_{\gamma^0}$, which is ${\cal B}^N_{\gamma^0}$,
the set of all four valent graphs that can be reached from
$\gamma^0$ in $N$ moves, but cannot be reached from
$\gamma^0$ by any path in fewer than $N$ moves.  
A pseudomanifold in ${\cal B}^N_{\gamma^0}$ will be
labeled $\gamma^N_{\alpha_1, ...\alpha_N}$ where, for example, 
$\gamma^2_{\alpha_1, \alpha_2}$ is the $\alpha_2$'th labeled
pseudomanifold
that can be reached from $\gamma^1_{\alpha_1}$.

It is also convenient to use the following terminology, borrowed from
considerations of combinatorial chemistry\cite{stu-ap}.  
We will call the set ${\cal S}^1_{\gamma^0}$ the {\it adjacent possible}
set of $\gamma_0$, as it consists of all the possible states that
could directly follow $\gamma_0$.  More generally, 
for any $N$, the set ${\cal B}^{N}_{\gamma^0}$
will be called the $N$'th {\it adjacent possible}, since it contains all the
possible new states available to the universe after $N$ steps that
were not available after $N-1$ steps.

It is clear that the for states composed of a large number of
labeled tetrahedra, the $N$'th adjacent possible sets grow
quickly, as is typical for combinatorial systems.

We may make some straightforward observations about the
sets ${\cal S}^N_{\gamma^0}$.

\begin{itemize}

\item{}Given two pseudomanifolds  $\alpha$ and $\beta$ 
in ${\cal S}^N_{\gamma^0}$, we will say that $\alpha$
generates $\beta$
if there is a single move that takes $\alpha$ to
$\beta$. (For example $\gamma^1_{\alpha_1}$
generates $\gamma^2_{\alpha_1, \alpha_2}$.)
${\cal S}^N_{\gamma^0}$ then has the structure of a supergraph
${\cal G}_{\gamma^0}^N$, which is a directed graph whose nodes
consist of the elements of ${\cal S}^N_{\gamma^0}$, connected by
directed edges that represent generation.  

\item{}A path $p$ in ${\cal S}^N_{\gamma^0}$ is a list
of  pseudomanifolds$\gamma_1,...\gamma_m$, each of whom generates
the next.  If there exists a path $p$ that runs from 
$\alpha$ to $\delta$, both elements of ${\cal S}^N_{\gamma^0}$
 then we may say that $\alpha \leq \delta$, or ``$\alpha$
precedes $\delta$".   ${\cal S}^N_{\gamma^0}$ thus has the
structure of a partially ordered set.  

\end{itemize}

There are corresponding statements for $\Omega_G$, the space
of all finite labeled pseudomanifolds.  We may define the set 
${\cal M}^N_{\gamma^0}$, an
element of which is a labeled pseudomanifold $\Gamma$.    This
corresponds to all elements of $\Omega_G$ which may be reached
in $N$ steps from an initial labeled pseudomanifold $\gamma_0$.
We may extend the
relations just defined to the elements of ${\cal M}^N_{\gamma^0}$.
Thus, given two labeled pseudomanifolds $\Gamma$ and $\Delta$, we may
say $\Gamma$ generates $\Delta$ if the graph $\gamma$
of $\Gamma$ generates the graph $\delta$ of $\Delta$, with the
obvious extensions to the notion of a path.  Thus, ${\cal 
M}^N_{\gamma^0}$
has as well the structure of a partially ordered set.  In
addition, we have the ``boundary" of ${\cal M}^N_{\gamma^0}$,
consisting of all the labelings of the elements of
${\cal B}^N_{\gamma^0}$, which we may call
${\cal A}^N_{\gamma^0}$.

We may note that neither ${\cal M}^N_{\gamma^0}$ nor
${\cal S}^N_{\gamma^0}$ are causal sets, as for $N$ large enough 
there will be closed
paths that may begin and end on a graph 
$\gamma \in {\cal S}^N_{\gamma^0}$.

We may note that there is an obvious map 
$r: \Omega_G \rightarrow \Omega$ in which labels are erased.

We consider  ${\cal M}^N_{\gamma^0}$ to be then the discrete analogue
of Wheelers superspace.  This is suggested by the fact that the
labeled pseudomanifolds diagonalize observables that measure the three
geometry.  We may note that just as in the
continuum case we may put a metric on 
${\cal M}^N_{\gamma^0}$.  If $\alpha > \beta$ or $\beta > \alpha$
then we may say that $\alpha$ and $\beta$ are causally related.
In this case, the metric $g(\alpha , \beta )=n$, the length of the
shortest path that connects them.    Thus, as in the continuum case,
the metric gives the superspaces a poset structure.

\section{Some simple models}

We will now illustrate some of the issues involved in the continuum
limit, using the frozen model as an example.
This model is similar to dynamical triangulation
models of Euclidean quantum gravity, but it differs from those
because of the role of the causal structure.  To write it down
more explicitly, we let the index $c$ take values over the
four types of causal structure: 
$c \in \{1 \rightarrow 4,  2 \rightarrow 3 ,
3 \rightarrow 2, 4 \rightarrow  1 \} $

There are then four amplitudes ${\cal A}[c]$ that must
be specified. We may write them in terms of amplitudes and
phases as,
\f
{\cal A}[c]= a_c e^{\imath \theta_c }
\ff
The amplitude for a history is then given by
\f
{\cal A}[{\cal M}]=\prod_c ({\cal A}[c])^{N_c}
\label{localamp}
\ff
where $N_c$ is the number of occurances of the $c$'th
causal structure in the history.  These of course satisfy
\f
N=\sum_c N_c  .
\ff

The model  has four parameters, which are the four
complex numbers ${\cal A}[c]$.  It can be further simplified so that 
it
depends only on fewer parameters.  One way to do this is to insist 
that
the amplitude are pure phases, so that all four moves have
equal probability, but with certain phases, 
\f
{\cal A}[c]=e^{\imath  \theta_c}
\ff
We can further simplify by insisting that each of the pair of moves 
that are
time reversals of each other have the same phase, 
this means that\footnote{The reader may wonder why we assign the 
time reversed amplitude to
be equal to the original, rather than its complex conjugate. The
answer is that we want a process followed, by its time reversal,
to be distinct from the process where nothing happens. In general
relativity a process and its time reversal are related by a 
diffeomorphism and thus have equal actions, thus in the quantum
theory they are given by equal amplitudes.}
\f
{\cal A}[1\rightarrow 4 ] ={\cal A}[1\rightarrow 4 ] = e^{\imath 
\alpha}
\ff
\f
{\cal A}[2 \rightarrow 3 ] ={\cal A}[3 \rightarrow 2 ] = e^{\imath 
\beta}
\ff

To write the amplitude let us then define 
$\lambda= {1 \over 2}(\alpha + \beta )$
and $\mu= {1 \over 2}(\alpha -  \beta )$. The total amplitude of 
a history $\cal M$ is then,
\f
{\cal A}[{\cal M}]= e^{\imath ( \lambda N_{total} + \mu N_{diff}  )}
\ff
where 
\f
N_{diff}=N[1 \rightarrow 4 ] + N[4 \rightarrow 1 ] - N[2 \rightarrow 
3 ] -
N[3\rightarrow 2 ] 
\ff

We see that as $N_{total}$ is proportional to the four volume, 
$\lambda$ plays the role of a cosmological constant.  It is 
interesting to compare this
to the action for dynamical triangulations, which is of the form
$S^{DT}= \lambda N_{total} + \kappa N_2$ where $N_2$ is the number of 
two simplices
which is a measure of  the averaged spacetime scalar curvature.  This 
suggests that if there is a continuum
limit $N_{diff}$ might also be a measure of the averaged spacetime 
curvature scalar,
suitable for spacetimes of Minkowskian signature.  

\section{The classical limit of the frozen models}

Of course, the actual behavior of the evolution described by
the theory will depend on
the details of the amplitudes ${\cal A}[c]$.  However, it is useful
to ask whether any conclusions can be drawn about the evolution in
the case that we have no information about the actual forms of the
amplitudes
Let us make the simplest possible assumption, which is that
all the amplitudes are given by some random real phase, so that
${\cal A}[c]= e^{i\theta}$.  Then the amplitude for any
path $p$ is $exp[i \theta n(p)]$.  

In this case we can draw some simple conclusions as
follows.  Consider the amplitudes 
${\cal A} [ \Gamma_0 \rightarrow \Gamma_f ]$ for all labeled 
pseudomanifolds
$\Gamma_f \in {\cal M}^N_{\gamma^0}$.  It is clear that
for $\Gamma_f \in  {\cal A}^N_{\gamma^0}$ the amplitudes
${\cal A} [ \Gamma_0 \rightarrow \Gamma_f ]= W e^{\imath \theta N}$,
where $W$ is the number of inequivalent ways to reach $\Gamma_f$ in
$N$ steps.  Thus, the amplitudes evolve in such a way that the 
amplitudes
for the states on the boundary is always a coherent phase.  

On the other hand, consider a $\Gamma_t$ which is
in the interior of ${\cal M}^N_{\gamma^0}$.  
Let this be an element that is in ${\cal A}^M_{\gamma^0}$
for some $M << N$.  
There will typically
be a number of different paths that reach $\Gamma_t$, with  a variety 
of
different path lengths.  The number of such paths will grow rapidly 
with
$N$, as long as $M << N$.  The total amplitudes for such labeled 
pseudomanifolds
to be reached after $N$ steps then will by 
${\cal A} [ \Gamma_0 \rightarrow \Gamma_f ] = 
\sum_{r} e^{\imath \theta r}$ with $r$ a finite set of
integers $ M \geq r \geq N$.  As $N$ grows large this set grows, 
and there are typically no interesting correlations amongst them.
In this case, as $N$ grows large then 
${\cal A} [ \Gamma_0 \rightarrow \Gamma_f ]  \approx 0$.

This means that for $N$ large, most of the amplitude predicted by
the path integral (\ref{pi}) with these assumptions will be concentrated on
${\cal A}^M_{\gamma^0}$ and a narrow shell trailing it.

This may be considered to be a form of the classical limit, because
as $N$ grows, the amplitude to have evolved from $\Gamma_0$ to
a state $\Gamma_f$ by an $N$ step path is concentrated on those
states that can be reached in $N$ steps, but no fewer.  This means 
that
as $N$ increases the amplitude is evolving along geodesics of the
metric $G$ defined in the discrete superspace.

\section{Dynamics including the parameters}

In the class of theories we have formulated here the dynamics of the 
theory
is given by four functions ${A}[c]$ which give the amplitude
for each four simplex which is added to a history as the result of
a Pachner move.  These functions depend on the  causal structure $c$ 
and labels on the 4-simplices.
By using the requirement that the functions are invariant under
permutations of the elements of the four simplex that do not change 
the
causal structure, we can reduce the functions $A[c,p]$
to particular forms which depend on a set of parameters, $p$,
which live in a parameter space $\cal P$.
The main dynamical problem is to find the set 
${\cal P}^* \subset {\cal P}$ 
such that the amplitudes defined by the sum over histories (\ref{pi}) 
has a good classical limit.

However there is clearly something unsatisfactory about this 
formulation.
No fundamental theory can be considered acceptable if it has a large 
number
of parameters which must be finely tuned to some special values in 
order
that the theory reproduces the gross features of our world.
Instead, we would prefer a theory in which the critical behavior 
necessary
for the existence of the classical limit was achieved automatically.
Theories of this kind are called ``self-organized critical systems".

One possibility is that the parameters $p$ which determine the 
amplitudes for the different evolution moves are themselves dynamical
variables which evolve during the course of the evolution of the 
system
to values which define a critical system with a good continuum limit.

Here is one form of such a theory.  We associate to each tetrahedron in
the model, ${\cal T}_i$ a value of the parameters $p_i$.  When a
move is made it involves $n < 5$ tetrahedra.  We will assume that
the amplitude of the move is given by $A[c,<p>]$ where
$<p>$ is the average of the $p_i$ among the $n$ tetrahedra involved
in the move.  The move creates $5-n$ new tetrahedra.  We assign to
each of them  the new parameters $<p>$.
This rule guarantees that those choices of parameters that spread
the most widely through the population of tetrahedra govern the
most amplitudes.  In this way, the system itself may discover and
select the parameters that lead to criticality, and hence a classical
limit.

Other rules for the new parameters may be contemplated.  Another 
choice is the following. The set of parameters $p_\alpha$ are
divided randomly into $n$ sets.  The new $p_\alpha$'s in each
of these sets are taken from the corresponding values in one of
the $n$ ``parent" tetrahedra that were input into the move.  This
distribution of the parameters is made separately for each of the
$5-n$ new tetrahedra.

The reader may object that the possibility for giving different rules 
for
the choices of parameters violates our intention that the system 
choose
its own laws.  However, this is not the point. There is no
way to avoid making a choice in giving rules to the system.   What we
want to avoid is the circumstance that the rules which result in a
classical limit are so unlikely that it seems a miracle that they be
chosen properly.   What would be more comfortable is an evolution rule
that has no sensitive dependence on a choice of parameters that 
results
in the system naturally having a classical limit.  By making the 
system
choose the parameters itself, on the basis of a rule that selects 
those
that lead to the most efficient propagation of information, we may
make it possible for the system to tune itself to criticality.

\section{Concluding remarks}

In this paper we have explored the suggestion\cite{fl1} that
the problem of the continuum limit of a certain class of quantum 
theories
of gravity may be explored through methods developed to study
non-equilibrium critical phenomena.   We have identified two key 
obstacles to the realization of this suggestion.
First, techniques must be developed which allow efficient computations
to be done on the class of theories described here. Second, those 
techniques must allow us to identify critical phenomena in real 
quantum
mechanical path integrals, in which one has a problem like quantum
directed percolation, in which histories are weighed by complex
amplitudes.  When these obstacles are overcome we will have the tools
we need to discover which non-perturbative quantum theories of gravity
have  good continuum limits which reproduce classical general 
relativity
interacting with quantized matter degrees of freedom, in a way that 
does
not rely assumptions about the applicability of Euclidean methods to
theories of dynamical causal structure, that are unjustified and
likely false.

\section*{ACKNOWLEDGEMENTS}

We are first of all thank Fotini Markopoulou, who was a coauthor of 
an  early version of this paper, for collaboration and suggestions. 
Conversations with Sameer Gupta has also been very helpful to 
formulating these ideas, and we are also endebted to him for comments
on a draft of this paper.
We also thank Jan Ambjorn, Kostas Anagastopolous, Per Bak,
Jayanth Banavar,  Roumen Borissov, Louis Crane, David Meyers, Maya Paczuski
and Carlo Rovelli
for helpful conversations about non-equilibrium critical phenomena and
evolving spin networks.  This work was supported by the National 
Science Foundation and NASA's astrobiology program and a
grant from the Jesse Phillips Foundation.

\end{document}